\documentclass[proof]{WileyASNA-v1}
\usepackage{color}

\articletype{Comment}%

\received{27 October 2021}
\revised{}
\accepted{}

\begin{document}

\title{Initial Energy of a Spatially Flat Universe -- a Hint of its Possible Origin}

\author[1]{Fulvio Melia*}

\authormark{Fulvio Melia}

\address[1]{\orgdiv{Departments of Physics and Astronomy, and the Applied Math Program}, 
\orgname{The University of Arizona}, \orgaddress{\state{Tucson, Arizona}, \country{U.S.A.}}}

\corres{*\email{fmelia@arizona.edu}}


\abstract{The evidence for a Big Bang origin of the Universe is truly compelling, though its cause
remains a complete mystery. As the cosmic spacetime is revealed to us with ever improving
detail, however, we are beginning to refine the range of its possible initial conditions---at
least within the framework of current physical theories. The Universe, it seems, is spatially
flat, and here we discuss in clear, straightforward terms why this trait implies a cosmos with
zero {\it kinetic} plus {\it gravitational} energy, though apparently not zero {\it total}
energy. Such an outcome has far reaching consequences because of the possibility that the
Universe may have begun its existence as a quantum fluctuation. Was this from `nothing,' or
perhaps a pre-existing vacuum? A non-zero total energy would seemingly preclude the former
scenario, but not necessarily the latter, though this would then raise the question of how a
fluctuation with non-zero energy could have lived long enough, or classicalized, for us to see
it 13.5 billion years later. The high-precision measurement of the Universe's spatial
curvature may thus constitute the first tangible piece of evidence impacting a possible
quantum beginning.}


\maketitle

\section{Introduction}
Over the past half century, many attempts have been made at `calculating' the energy of
the Universe using a variety of approaches with a diverse set of underlying assumptions and
definitions. With very few exceptions, these studies have inevitably concluded that
the Universe has zero total energy. Given that the definition of what constitutes a
viable contribution to this energy has not been uniformly consistent, however, one may
question whether all of these treatments can be correct. In fact, with such ambiguity, how 
can one be certain that any of them are physically meaningful? There is also the suspicion
that `zero energy' is a fundamentally preferred answer because such an initial condition 
lends considerable support to the notion that the Universe may have started as a quantum 
event from `nothing' or, at the very least, from a pre-existing vacuum. These two scenarios 
are quite different, of course, though the condition of zero energy seems to be a favoured 
requirement in both cases. 

The question of how much energy the Universe possesses first started emerging via a topological 
argument due to P. Bergmann quoted by \cite{Tryon:1973}, and then elaborated upon by several 
other authors, including \cite{Albrow:1973}, \cite{Guth:1981}, \cite{Rosen:1994},
\cite{Cooperstock:1994}, and several more since then. All along, an important 
(though not exclusive) motivation for addressing this question has been the proposal first 
made by \cite{Tryon:1973} that our Universe is a fluctuation of the vacuum, to be 
understood in the context of quantum field theory. Tryon conjectured that the appearance 
of the Universe `from nowhere' would require it to have a net zero value for all its conserved 
quantities, the most important of which is energy. He argued that the enormous amount of mass 
energy in the cosmos is canceled by its (negative) gravitational potential energy, and 
proceeded to show that these two quantities, contained within the Hubble radius 
$R_{\rm h}\equiv c/H$, where $H$ is the Hubble parameter, differ from each other only
by a factor of order unity.

Tryon made no mention of the positive energy due to the Universe's expansion, however, and 
envisaged the initial fluctuation as having taken place in the vacuum of a larger space in 
which the cosmos is embedded. As such, the total fluctuation energy $\Delta E$ need not have 
been zero, though one would then have to explain, via the Heisenberg uncertainty principle, 
$\Delta E\Delta t\sim \hbar/2$, how the Universe could have lived long enough for us to see it 
$\Delta t\sim 13.5$ billion years later. Tryon's paper was quite speculative, but it was 
the seminal work that introduced a quantum origin as a possible explanation for the Big Bang. 
He even went so far as to point out that a pre-existing vacuum should be unstable against 
large-scale fluctuations in the presence of a long-range, negative-energy, universal 
interaction, such as gravity.

To illustrate how broadly this discourse has meandered over the years, it has even been
suggested that a pre-existing vacuum would have permitted an enormous range of fluctuations
\citep{Sekrst:2020}, but intelligent beings would see only a Universe whose outcome is suitable 
for them to exist. This philosophically interesting argument, known as the anthropic
principle \citep{Barrow:1988}, is quite forgiving because only a Universe capable of
supporting life would be seen, no matter how statistically improbable it would be
compared to other possibilities. But notice that implicit in this argument is the
requirement that there be a distribution of possible quantum fluctuations, among which
the one that produced our highly improbable Universe was just a representative of 
many (perhaps an infinite number of) others.

This was also an issue considered in some detail by \cite{Ellis:2006}, who
argued that a choice between various contingent initial conditions somehow occurred at
the Big Bang, but it is not clear how that selection came about, at least not in the
absence of an anthropic principle. The questions one may ask in this context seem 
endless, even beyond scientific exploration. For example, why did the Universe's
expansion as a bubble from that pre-existing vacuum  start when it did, rather than
at some previous time in a presumably pre-existent eternity? Attempts at avoiding such
difficult or unanswerable queries often avoid a beginning by subscribing to cyclic 
phases, as in Tolman's series of expansion and collapse cycles \citep{Tolman:1931},
or Linde's eternal chaotic inflation \citep{Linde:2007} and Khoury et al.'s ekpyrotic 
universe \citep{Khoury:2001}.

Several alternative proposals have also been made for initiation events in the absence 
of space and time, though the distinction between such models, in which literally `nothing'
existed prior to the Big Bang and those with a pre-existing vacuum, is sometimes blurred. 
The obvious corollary to the former is that both space and time would have been created 
at the Big Bang along with the Universe itself. This class of cosmological origins 
includes the `Universe from Nothing' concept promoted by several workers, such as 
\cite{Krauss:2012}, \cite{Vilenkin:1982}, \cite{Zeldovich:1984} and, more recently, 
\cite{He:2014}.  But this approach tends to face greater scrutiny than the pre-existing 
vacuum idea, for the simple reason that an `ontic seed' must have been necessary 
to actually cause the fluctuation, as has been argued by \cite{Isham:1994}.
In addition, this class of models is distinguished from the pre-existing vacuum
idea by the absolute requirement of zero energy throughout the Universe's history. 

For such reasons, the `Universe from Nothing' proposals certainly receive more vigorous 
criticism than their pre-existing vacuum counterparts. For example, \cite{Kohli:2016} 
describes several flaws in the fundamental basis of these models, amplifying the point made 
earlier by Isham. On the one hand, he argues, these models assume that the Universe originated 
from nothing, yet they require all the complex machinery of variational principles, 
differential and pseudo-Riemannian geometry, topology, general relativity and quantum 
field theory, without ever addressing the question of where these come from. Indeed, 
even the theory used to describe the quantum event itself often requires an underlying 
superspace, which is absent by default. 

Having said this, our goal in this paper is not so much to argue for or against either class 
of models, but merely to motivate our exploration of what the recent cosmological observations 
are telling us about the most critical ingredient in all of these scenarios, i.e., the total 
energy ($E$) of the Universe. As we saw above, the first attempts (by Bergmann and Tryon) were
rather simple, and probably unrealistic. For example, Bergmann's closed universe would not
be consistent with the spatial flatness we are measuring today. And, as noted earlier, Tryon
did not actually demonstrate that $E$ is exactly zero, partly because he did not include all 
possible contributions to it. But these were only the first forays; there have been many 
others since then.

Nevertheless, attempts at calculating the energy of the Universe prior to the 2000's were largely
based on the use of pseudotensors, rendering those approaches highly questionable \citep{Faraoni:2003}.
Pseudotensors may be useful tools in studies of bounded systems that are asymptotically
Minkowskian, but the Universe is not bounded for Friedmann-Lema{\^i}tre-Robertson-Walker 
(FLRW) metrics with a spatial curvature constant $k\le 0$ (see Eq.~\ref{eq:metric} below). 
A more reasonable proposal for calculating $E$ with a full general relativistic treatment 
was introduced by \cite{Faraoni:2003}, who showed that $E$ is 
constant for $k\le 0$. They then argued that these universes are asymptotically Minkowskian 
and must therefore have a vanishing energy, concluding that $E=0$ from the beginning. 

But it is not clear that this argument works for an FLRW spacetime with $k\not=0$, since it 
is not asymptotically Minkowskian when expressed in terms of comoving coordinates, though 
one can demonstrate that it is equivalent to Minkowski space with an appropriate gauge 
transformation \citep{Melia:2012red}. Thus, one may conclude for $k\not=0$ that $E$ is 
constant, but not that it is also zero. For example, \cite{Kohli:2016} points out 
that any spatially  homogeneous and non-static universe, i.e., one that does not contain 
a global timelike Killing vector, is necessarily {\it not} asymptotically flat. On the 
other hand, their result for $k=0$ appears to agree with one of our conclusions in this 
paper, and we shall return to this in \S~IV below. It should be pointed out, though, 
that Faraoni and Cooperstock did not include all possible contributions to $E$. They 
based their approach on the use of a Lagrangian, from which they derived the Hamiltonian, 
so they certainly considered both the `kinetic' and `potential' energy components, 
but did not include the cosmic fluid's energy density $\rho$ itself in the 
overall energy budget. 

In this paper, we shall take a different---much more pedagogical---approach to interpret
the value of $E$, based on the observational determination that our Universe is 
spatially flat (i.e., $k=0$). We shall focus our attention on the most reliable `energy 
conservation' equation we have for the FLRW spacetime, i.e., the Friedmann equation (see 
Eq.~\ref{eq:Friedmann} below). We shall interpret its meaning using a Newtonian approach, 
though strongly motivated by the Birkhoff-Jebsen theorem \citep{Birkhoff:1923} and its corollary. 
And we shall clearly distinguish between the different forms of energy contributing to $E$.

\section{The Cosmic Spacetime}
Theoretical cosmology is founded on the Cosmological Principle (CP), comprising two 
essential symmetries: homogeneity and isotropy, at least on spatial scales \citep{Yadav:2010} 
larger than $\sim 300$ Mpc. An important consequence of this principle is the highly simplified 
form of the (FLRW) metric used to describe the cosmic spacetime \citep{Melia:2020}, normally expressed as
\begin{equation}
ds^2 = c^2 dt^2 - a^2(t)[dr^2 (1 - kr^2)^{-1} + r^2(d\theta^2 + \sin^2\theta d\phi^2)]\;.\label{eq:metric}
\end{equation}
This form is written in terms of comoving coordinates, in which $t$ is the proper (sometimes
called `cosmic') time, $r$ is the radius, $\theta$ is the poloidal angle and $\phi$ is the azimuth.
Unless an object in the Universe exhibits so-called `peculiar' motion, its comoving coordinates remain
fixed. Proper motion and the universal expansion itself are instead given in terms of the proper radius,
$R=a(t)r$, in which the expansion factor $a(t)$ is solely a function of time, not position. The 
most important element in this equation for this paper is the spatial curvature 
constant $k$, whose value is $+1$ for a closed universe, $0$ for a flat, open universe, and $-1$ 
for an open universe. In reality, $k$ could be any real number, though one typically renormalizes 
the radius $r$ in any given cosmological model in order to reduce it to one of these three 
distinct integer values. 

If one accepts the FLRW metric as the basis of our cosmology, most of the work in building a
framework for the origin and evolution of the Universe revolves around the behavior of $a(t)$. 
Each of the models we have today assumes some combination of matter and energy, predicting its 
own distinct time dependence of the expansion factor. And in the majority of cases, the 
underlying dynamical equations for $a(t)$ are obtained by introducing the metric in 
Equation~(\ref{eq:metric}) into Einstein's field equations of general relativity \citep{Melia:2020}. 
The one most relevant to us here is the Friedmann equation,
\begin{equation}
H^2\equiv\left({\dot a\over a}\right)^2={8\pi G\over 3c^2}\rho-{kc^2\over a^2}\;,\label{eq:Friedmann}
\end{equation}
in which an overdot denotes a derivative with respect to $t$, $\rho$ represents the total 
energy density in the cosmos, and $H\equiv \dot{a}/a$ is the Hubble parameter. In preparation
for our discussion in \S~IV, we emphasize here that the Friedmann equation is exact. No
approximations were made in deriving it directly from the full expression of Einstein's
equations in general relativity.

\section{The Flatness Problem}
Over the past several decades, precision measurements \citep{Spergel:2003,Planck:2016} of the cosmic 
microwave background (CMB) have shown that the Universe is very nearly flat, perhaps completely flat, 
with an inferred spatial curvature constant $k\approx 0$. Therefore, the density $\rho$ appears to be
at (or very near) the so-called `critical' density
\begin{equation}
\rho_c\equiv {3c^2H^2\over 8\pi G}\;,\label{eq:rhoc}
\end{equation}
obtained by setting $k=0$ in Equation~(\ref{eq:Friedmann}).

A density close to $\rho_c$ can be difficult to explain, however, unless $k$ has in fact always
been exactly zero. To see why, let us define
\begin{equation}
\Omega\equiv \rho/\rho_c\;,
\end{equation}
and re-write Equation~(\ref{eq:Friedmann}) as follows:
\begin{equation}
1=\Omega(t)-{kc^2\over {\dot{a}^2}}\;.
\end{equation}
Using subscript `0' to denote quantities pertaining to the present time $t_0$, we therefore find that
\begin{equation}
\Omega(t)-1=\left({\dot{a}_0\over\dot{a}}\right)^2\left(\Omega_0-1\right)\;.\label{eq:flat}
\end{equation}
This equation may not look very peculiar, but let's focus on the multiplicative factor 
$(\dot{a}_0/\dot{a})^2$ on the right-hand side. Today, we've all become accustomed to the idea 
that the Universe may be accelerating, but this was not the situation at early times; it was 
actually decelerating at an enormous rate. Let's say for simplicity that $\rho$ was dominated by 
radiation. This would certainly have been the case during the first few hundred thousand years, 
though not afterwards. One may then put $\rho\propto a^{-4}$ in Equation~(\ref{eq:Friedmann}), 
since the energy density of a radiation field scales inversely with the spatial volume 
($\propto a^3$), and inversely with the wavelength ($\propto a$), due to redshifting as the 
Universe expands. One can therefore easily see from the solution to Equation~(\ref{eq:Friedmann}) 
that $a(t)\propto t^{1/2}$, as long as $k$ is small. (Incidentally, had we assumed 
$\rho$ was dominated by matter instead, we would have found that $a(t)\propto t^{2/3}$, but 
such differences have no impact on this argument.) The prefactor in Equation~(\ref{eq:flat}) 
would thus be $({\dot{a}_0/\dot{a}})^2\sim t/t_0$. So comparing the quantity $\Omega(t)-1$ 
at the GUT (Grand Unification Theory) time ($t_{\rm GUT}\sim 10^{-35}$ seconds; see below) 
with its counterpart 13.5 billion years later, we see that it must have been $\sim 10^{48}$ 
times smaller compared to its value today. Any indication that $\Omega_0\approx 1$ today 
therefore suggests that $\Omega(t)$ must have been {\it fine tuned} to nearly exactly $1$ 
as $t\rightarrow 0$. 

This fine-tuning is the origin of the so-called `flatness' problem in cosmology, in the sense 
that it seems to require yet one more special initial condition without any physical 
explanation \citep{Lightman:1990,Hawking:1974}.  But this clearly stems from the view that the Universe 
could have been born with any value of $k$, so an initial condition $\Omega(t)=1$ with 
$k\not=0$ would appear to be highly unlikely. To address this issue, several different 
approaches have been taken to explain why the Universe appears to be flat (or nearly flat) 
today, in spite of it possibly having begun with an arbitrary spatial curvature.

One possible solution is to again invoke the anthropic principle \citep{Collins:1973}, arguing 
that only a Universe with the correct density to form galaxies and stars would give rise to
intelligent observers who would then inquire about the initial conditions. It should then
not be surprising that we do not live in other types of universes with unsuitable values
of $\Omega(t)$ at the beginning.

But the explanation that currently enjoys the most widespread acceptance (if not outright belief)
is inflation, a period of exponential growth in the Universe only $\sim 10^{-35}$ seconds
after the Big Bang, possibly associated with the separation of the strong and electroweak
forces at the aforementioned GUT scale. In his paper, 
\cite{Guth:1981} proposed that such an early phase of very rapid expansion would solve
several puzzles in cosmology, including the flatness problem. A requirement of inflation
is that the density $\rho$ remain constant for a brief, though critical, period, and it 
is not difficult to see from Equation~(\ref{eq:Friedmann}) that, under such circumstances,
$a(t)$ would grow exponentially (again ignoring any small value of $k$). Thus, any value
of $\Omega(t)-1$ at the beginning would rapidly shrink to near zero by the end of inflation, 
due to the enormously large factor $\dot{a}\sim e^{Ht}$ in the r.h.s. denominator of 
Equation~(\ref{eq:flat}). And the Universe would emerge back into the standard hot Big
Bang configuration to resume its expansion resulting in $\Omega_0\sim 1$ today. In other
words, inflation could in principle wash out any initial condition associated with
$\Omega(t)$ resulting from an arbitrary value of $k$.

Inflationary theory is not universally accepted, however, and recent observations are starting
to raise doubt about whether it can actually explain the various horizon problems plaguing
the standard model \citep{Melia:2018}, and the fluctuation spectrum seen in the CMB 
\citep{Liu:2020}. There are still too many gaps in the theory, and the continued refinement
of cosmological measurements may eventually disprove it. For example, even after four decades
of development, inflationary cosmology still lacks specificity regarding the field that
is driving it \citep{Bird:2008}. The various versions of the theory spawned by this ambiguity
contain parameters and initial conditions that themselves require fine-tuning \citep{Ijjas:2014}, 
analogously to how `unnatural' the initial density would be in the absence of inflation 
(see Eq.~\ref{eq:flat}).

There are therefore good reasons to explore other explanations for the flatness of the
Universe, even more so today because the ever improving observations are telling us that
$k$ is not only close to zero, but is probably exactly zero. So the `flatness' problem
may be moot. It may not be a problem at all, if it turns out that there is a physical
reason why the Universe had to have perfect spatial flatness as one of its initial conditions.

\section{Cosmic Energy}
Today, we have a better understanding of what $k$ stands for. Spatial curvature is not merely 
a geometric property of the spacetime---it results from a well defined physical attribute,
which can be understood with a more thorough inspection of the Friedmann Equation~(\ref{eq:Friedmann}).
Though this expression derives from Einstein's field equations, it is not difficult to find a simple 
heuristic argument for its physical meaning, which includes an explanation for the spatial curvature 
constant $k$. This is most easily done by invoking the corollary to the Birkhoff-Jebsen theorem 
\citep{Birkhoff:1923}, which allows us to understand why the influence of any {\it isotropic}, 
external source of gravity completely cancels within a spherical cavity. This theorem is a 
relativistic generalization of Newton's theory---that the gravitational field outside a spherically 
symmetric body is indistinguishable from that of the same mass concentrated at its center. It was 
first pointed out by \cite{Weinberg:1972} that, due to this symmetry, a limited use of 
Newtonian gravity is permitted for some cosmological applications, such as the one we are 
concerned with here. 

A consequence of isotropy is that every observer in the cosmos experiences {\it net} zero
acceleration due to the surrounding mass. But in fact the {\it relative} 
acceleration between any two points is not zero; it depends on the mass/energy content 
between them. This is the reason why the corollary to the Birkhoff-Jebsen theorem is so 
germane to our discussion, because it describes the field {\it inside} an empty spherical 
cavity at the center of an isotropic distribution, which may be used to calculate the
relative acceleration locally. The metric inside such a cavity is Minkowskian, a situation 
not unlike we find in electromagnetism, where we would calculate the electric field inside 
a spherical cavity embedded within an otherwise uniform charge distribution to be zero. So 
even in the classical limit, it is safe to argue that the medium exterior to a sphere can 
be represented as a sequence of shells with an ever increasing radius, each of which 
produces zero effect within the cavity.

\begin{figure}
\vskip 1cm
\centerline{
\includegraphics[angle=0,scale=0.7]{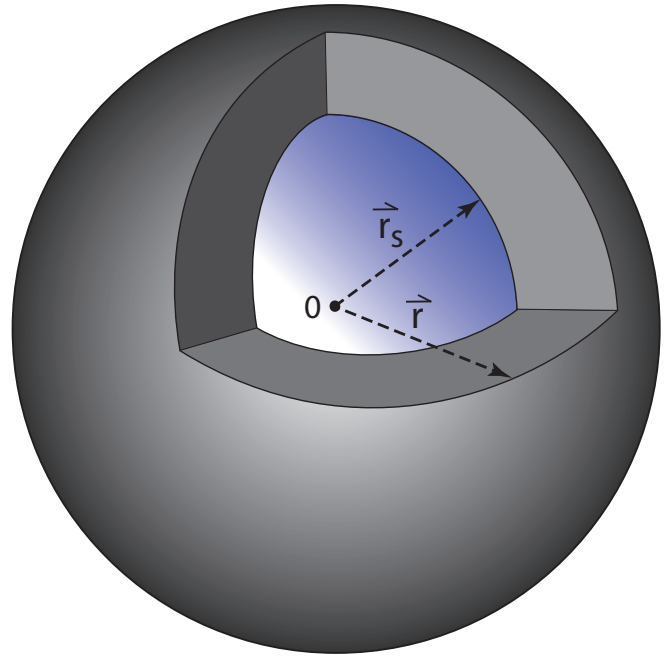}}
\vskip 0.2in
\caption{The Birkhoff-Jebsen theorem \citep{Birkhoff:1923} and its corollary tell us that in 
an isotropic Universe, the mass-energy everywhere outside (gray) of a spherical region (blue)
of radius $r_s$ (i.e., at all $r>r_s$) has no gravitational influence at $r\le r_s$. Thus, 
the gravitational acceleration of any point on the shell at $r_s$ {\it relative to an 
observer} at the origin depends solely on the mass-energy contained within this (blue) 
sphere.}\label{fig1}
\end{figure}

Let us imagine `carving' a sphere of radius $R_{\rm s}(t)=a(t)r_{\rm s}$ out of an otherwise 
homogeneous, isotropic universal medium (fig.~\ref{fig1}).  The rest of the Universe has 
zero gravitational influence inside this cavity. We then fill the sphere with density $\rho$ all 
the way up to an infinitesimal distance away from $R_{\rm s}(t)$. According to the Cosmological 
Principle, this density needs to be a function solely of $t$. In addition, the FLRW metric tells 
us that every pair of points within and outside the sphere recede from each other at a rate 
proportional to the expansion factor $a(t)$, which itself is the same everywhere. But according 
to the Birkhoff-Jebsen theorem and its corollary, we only need to consider the source of gravity 
within $R_s$ to determine the local dynamics of this region (i.e., the behavior of $a$) 
extending out to this radius. 

We place an observer at the center of this sphere, where they calculate the kinetic energy 
of a shell with thickness $dR$ at $R$ to be 
\begin{equation}
dK=4\pi R^2\,dR\,{\rho(t)\dot{R}^2\over 2c^2}\;,
\end{equation}
where the factor $c^2$ converts the energy density to a mass density. One
must be aware, of course, that there are limitations to how far one can take this
classical, heuristic approach. Inferring a mass density using this method for
the purpose of calculating a "kinetic" energy is reasonable for a matter dominated
Universe, perhaps even a matter plus radiation Universe, as long as the radiation
in the latter is `trapped' within the matter, allowing one to use the $E=mc^2$
conversion to infer the gravitational mass. This approach would not be sensible
for a purely radiation-dominated Universe, however. If the observer now
integrates this quantity from $r=0$ (their location) to $r=r_{\rm s}$ (the edge of the
sphere), they find that the total kinetic energy of the sphere relative to them is 
\begin{equation}
K={2\pi\over 5}{\rho(t)\over c^2}a^3{\dot{a}}^2\,{r_{\rm s}}^5\;.\label{eq:kinen}
\end{equation}

This sphere also has gravitational energy (remember this is a classical derivation). 
The observer finds that the corresponding potential energy of a shell at $R$ is
\begin{equation}
dU=-4\pi R^2\,dR\,{GM(R)\rho(t)\over c^2R}\;,
\end{equation}
where
\begin{equation}
M(R)={4\pi\over 3}\,R^3\,{\rho(t)\over c^2}
\end{equation}
is the total mass enclosed within this radius. Thus, after integrating this quantity from 
$r=0$ to $r_{\rm s}$, they find that the total potential energy of the sphere (relative
to them at the origin) is
\begin{equation}
U=-{16\pi^2G\over 15}{\rho(t)^2\over c^4}a^5{r_{\rm s}}^5\;.
\end{equation}

Since they are doing this calculation classically, they therefore conclude that the sphere
expanding away from them has a net {\it kinetic} plus {\it gravitational} energy 
\begin{equation}
E={2\pi\over 5}{\rho(t)\over c^2}a^3{\dot{a}}^2\,{r_{\rm s}}^5-
{16\pi^2G\over 15}{\rho(t)^2\over c^4}a^5{r_{\rm s}}^5\;,
\end{equation}
and if they re-arrange this slightly, a very recognizable form appears:
\begin{equation}
\left({\dot{a}\over a}\right)^2={8\pi G\over 3c^2}\rho(t)+
{5c^2E\over 2\pi \rho(t)\,a^5\,{r_{\rm s}}^5}\;.\label{eq:totalen}
\end{equation}
The local conservation of energy relative to the observer at the origin is thus simply
the Friedmann Equation~(\ref{eq:Friedmann}), so long as one interprets the spatial 
curvature constant to be 
\begin{equation}
k\equiv -{10\over 3\,{r_{\rm s}}^2}\left({\epsilon\over\rho}\right)\;,
\end{equation}
written in terms of the local kinetic plus gravitational energy density
\begin{equation}
\epsilon\equiv {3E\over 4\pi\,{R_{\rm s}}^3}\;,
\end{equation}
and the energy density $\rho$ in the cosmic fluid. {\it But note that $\epsilon$ does not
include $\rho$ itself}. Incidentally, calling the quantity $k$ defined in
Equation~(14) a constant would be justified only as long as $\epsilon$ and $\rho$ scale
in proportion to each other. In a classical context, this simply means that the sum
of kinetic plus gravitational energy (i.e., $E$) is conserved, which is what one would
expect without an energy non-conserving influence affecting the cosmic expansion. One
must be careful in general relativity, however, since the energy measured by an observer
can change in a time-dependent spacetime. As we shall discuss shortly, however, even this 
distinction may not matter for our interpretation of $k$ if $\epsilon$ is strictly zero, 
because then $\epsilon$ would always be zero in both the classical and relativistic contexts.

We thus see that a Universe with positive spatial curvature ($k>0$) contains negative net
energy ($\epsilon<0$) and is therefore bounded---for which the more common terminology is 
that the Universe is `closed'. Negative spatial curvature arises when this energy is positive 
($\epsilon>0$), which characterizes an unbounded (or `open') Universe. A Universe with zero 
net energy ($\epsilon=0$) is spatially flat ($k=0$), but also `open'. To put this in context, 
let us consider the following simple analogy. Imagine throwing a ball vertically into the air, 
ignoring any resistance. Its motion will depend on how its initial speed ($v_{\rm init}$)
compares to the value it needs ($v_{\rm esc}\equiv \sqrt{2GM_{\rm Earth}/R_{\rm Earth}}$) 
to escape from Earth. If $v_{\rm init}<v_{\rm esc}$, the ball slows down, stops and 
then falls back to its point of origin. The ball is bounded to Earth because its total 
energy is negative. If $v_{\rm init}>v_{\rm esc}$, the ball has positive energy and is 
unbounded. It will continue receding to infinity with a speed always greater than zero. 
The critical initial condition is $v_{\rm init}=v_{\rm esc}$, for which the ball rises 
indefinitely, though with an ever decreasing speed, coming to rest at infinity.

A spatially flat Universe therefore has zero local energy density $\epsilon$. Normally,
attempts at quantitatively handling gravitational energy in general relativity face the daunting
problem of localization \citep{Landau:1962}. Here, however, the $\epsilon=0$ condition holds 
at every point in the Universe, so it follows that the total kinetic plus gravitational
energy must also be zero, no matter how one chooses to do the integration.  This simplified 
heuristic argument therefore teaches us that spatial flatness is a unique, fundamental 
physical property of the Universe, which apparently contains equal portions of positive 
kinetic (expansion) energy and negative gravitational energy. Moreover, $k$ is constant, 
so this perfect balance must have been present from the beginning, constituting an important 
initial condition at the time of the Big Bang.

\section{Discussion}
It is tempting to think that perhaps our Newtonian approach may lack certain nuances 
manifested solely via the full general relativistic treatment. But remember that our 
calculation of $\epsilon$ has been carried out strictly on a local basis, with full 
justification from the Birkhoff-Jensen theorem. In fact, the proper radius of the 
sphere used in this construction, i.e., $a(t)r_s$, may be chosen arbitrarily small 
compared to the Hubble radius $R_{\rm h}=c/H$, because we are interested merely in 
the local value of $\epsilon$, point by point. And once $\epsilon$ is known for a 
given observer, every spacetime point in the FLRW cosmos has precisely the same value, 
at least according to the Cosmological principle. There is therefore nothing missing from 
Equation~(\ref{eq:totalen}) that would otherwise emerge from an analogous calculation using 
Einstein's equations. In particular, notice that this equation matches precisely the actual 
Friedmann equation (\ref{eq:Friedmann}) derived from general relativity.

The key result of this derivation is a clear, unambiguous identification of the energy $E$ in 
Equation~(\ref{eq:totalen}), which appears to be directly proportional to the spatial curvature 
constant $k$. It includes only kinetic plus gravitational energy, but not a contribution 
from $\rho$ itself. Our result therefore agrees completely with that of \cite{Faraoni:2003} 
in the case of $k=0$, who used an entirely different
method of identifying the kinetic and potential energy of the cosmic expansion. In
both cases, the conclusion is that $E$ is zero, but it only represents the kinetic and
gravitational energy. Thus, if the latest observations are telling us that $k=0$, we 
cannot avoid the conclusion that the Universe has {\it net positive} energy, when 
the integrated value of $\rho$ is taken into account.

Nevertheless, one must still contend with several possible caveats to this result. First, it
may turn out that $k$ is close to, but not exactly, zero, in which case there may still 
be some wiggle room to keep the overall sum of energies zero. We note, in this regard,
that the most precise measurement of the spatial curvature to date, from the {\it Planck} mission
\citep{Planck:2020}, is quoted as $\Omega_k=0.001\pm0.002$, where $\Omega_k$ is defined
as the last term in Equation~(\ref{eq:Friedmann}) divided by $H^2$. There is no question
this parameter is fully consistent with zero, but it may still allow for a very slight 
positive or negative bias. This condition is the essential caveat with this
work, for if it turns out that $k$ is in fact not exactly zero, then the conclusions
we draw here become moot. We note, in this regard, that observations other than those
pertaining to the CMB also indicate a likely flat Universe, consistent with {\it Planck}
\citep{Wei:2020a,Wei:2020b}, though the precision of these alternative
measurements is inferior to that of the CMB satellite missions. Nevertheless, one must
keep in mind that any discussion relating the total energy of the Universe to its
spatial flatness becomes much more difficult to resolve if $k$ is not zero, given in
part by the complications discussed earlier of handling the energy in a general
relativistic framework when the spacetime curvature changes with time. 

Second, inflation may have happened after all, in which case $\Omega_0-1$
in Equation~(\ref{eq:flat}) is close to zero today only because of the significant dilution
that took place during the inflationary expansion at $t\sim 10^{-35}$ sec. Having
said this, it is also possible that $k=0$ was an initial condition regardless of whether
inflation happened or not, in which case our inference on the total energy would be valid
nonetheless. It is beyond the scope of the present paper to discuss inflation at
greater length, but it might be helpful to point out that the observational evidence
in favor of such an event having occurred in the early Universe is rather slim. As noted 
earlier, the latest analysis of the {\it Planck} data suggests that a slow-roll inflaton 
potential could not simultaneously have solved both the CMB temperature horizon problem and 
provided a mechanism for seeding the primordial fluctuation spectrum \cite{Liu:2020}. And at 
a more fundamental level, it is even debatable whether these quantum fluctuations could have
classicalized to produce the classical large-scale structure we see today \cite{Melia:2021}.

Third, in addition to inflation, there are other possible reasons why the Universe
may appear close to spatial flatness today, even if $k=0$ was not an initial condition.
For example, a key factor in many `cyclic' models of the Universe is a so-called
ekpyrotic phase, in which the contraction preceding the Big Bang evolves ultra-slowly
\citep{Raveendran:2019,Ijjas:2019}. In such models, it is this ekpyrotic phase that can 
explain the smoothness and flatness of the Universe on large scales, and it can also
generate a distinctive primordial spectrum of scalar and tensor fluctuations.

So there are still many unknowns, and while our conclusions in this paper are suggestive, we 
would not call them `final'. We are beginning to get some guidance from the data concerning 
the nature of the Big Bang, but there is still much work to do before we can identify its 
true physical origin. 

\section{Conclusion}
In a way, this result completes the derivation begun by Tryon in his seminal paper \citep{Tryon:1973},
in which he attempted to balance the gravitational energy by the rest mass energy in the cosmic
fluid. As we now understand it, the kinetic energy of expansion exactly balances its 
gravitational counterpart, leaving $\rho$ as an `excess' positive contribution to $E$. 

If these results withstand the test of time, it would not be possible to reconcile the presently 
`measured' non-zero energy of the Universe with the requirements of a quantum birth from `nothing'. 
Perhaps a quantum fluctuation in a pre-existing vacuum might still be consistent, but one would 
need to understand how its energy $E$---in the context of our Universe's measured value of the Planck 
constant, $\hbar$---could have allowed it to grow and classicalize and live long enough for us to 
see it today. Would this be an indication that $\hbar$ had to be different prior to the Big Bang?


\subsection*{Author contributions}

Fulvio Melia is the sole author of this paper.

\subsection*{Financial disclosure}

None reported.

\subsection*{Conflict of interest}

The authors declare no potential conflict of interests.

\bibliography{ms}%

\vfill\newpage
\section*{Author Biography}

\begin{biography}{}{\textbf{Fulvio Melia} 
is an internationally recognized leader in astrophysics and cosmology, having previously published 
seven books in physics and astronomy, and approximately 400 refereed publications in the these 
fields. He has also served as associate editor of the Astrophysical Journal and Astrophysical 
Journal Letters.} 
\end{biography}

\end{document}